\documentclass{iau_FM}
\usepackage{graphicx}

\usepackage{color}

\title[The High-energy emission of jetted AGN] %% give here short title %%
{The High-energy emission of jetted AGN} %% give here the full title %%

\author[Daniel A. Schwartz]   %% give here short author list %%
{Daniel A. Schwartz$^1$}
\affiliation{$^1$Smithsonian Astrophysical Observatory, \\ 60 Garden
  St.,Cambridge, MA 02138, USA\\ email: {\tt das@cfa.harvard.edu}}
\pubyear{2018}
\jname{Astronomy in Focus, Volume 1} 
\editors{Volker Beckmann \& Claudio Ricci, eds.}
\newcommand{\axaf}{\mbox{\em Chandra\/}}
\newcommand{\arcsec}{\hbox{$^{\prime\prime}$}}
\newcommand{\farcs}{\mbox{$.\!\!^{\prime\prime}$}}

\newcommand{\degs}{\mbox{$^{\circ}$}}

\definecolor{darkgreen}{rgb}{.0, .6, .0}

\begin{document}
\maketitle

\begin{abstract}
  Quasars with flat radio spectra and one-sided, arc-second scale,
  $\approx$ 100 mJy GHz radio jets
  are found to have similar scale X-ray jets in about 60\% of such
  objects, even in short 5 to 10 ks \axaf\ observations. Jets emit in
  the GHz band via synchrotron radiation, as known from polarization
  measurements.  The X-ray emission is explained most simply, i.e.
  with the fewest additional parameters, as inverse Compton (iC)
  scattering of cosmic microwave background (CMB) photons by the
  relativistic electrons in the jet. With physics based assumptions,
  one can estimate enthalpy fluxes upwards of 10$^{46}$ erg s$^{-1}$,
  sufficient to reverse cooling flows in clusters of galaxies, and
  play a significant role in the feedback process which correlates the
  masses of black holes and their host galaxy bulges. On a
  quasar-by-quasar basis, we can show that the total energy to power
  these jets can be supplied by the rotational energy of black holes
  with spin parameters as low as $a=0.3$. For a few bright jets at
  redshifts less than 1, the \emph{Fermi} gamma ray observatory shows
  upper limits at 10 Gev which fall below the fluxes predicted by the
  iC/CMB mechanism, proving the existence of  multiple  relativistic
  particle populations. At large redshifts, the CMB energy
  density is enhanced by a factor (1+z)${^4}$, so that iC/CMB must be
  the dominant mechanism for relativistic jets unless their rest frame
  magnetic field strength is hundreds of micro-Gauss. 

%% add here a maximum of 10 keywords, to be taken form the file <Keywords.tex>
\end{abstract}

\firstsection % if your document starts with a section,
\section{Introduction}

Radio astronomy is intimately related to high energy astronomy.
Estimates of magnetic fields in the lobes of radio sources led to the
recognition that TeV electrons must be present to radiate via the
synchrotron mechanism. This motivated early suggestions
(\cite{Morrison58,Savedoff59}) that direct detection of celestial gamma rays might
be feasible.
%detect other manifestations of these electrons. 
Estimates of the energy contents of these radio lobes were extremely
large, posing difficulties for explaining the origin and possible
relation to the associated galaxy or quasar. The dilemma was solved by
theoretical explanations of how collimated beams of particles and
fields, i.e.  jets, could carry energy to the lobes (\cite
{Rees71,Longair73,Scheuer74,Blandford74, Begelman84}), and by direct
imaging of these radio jets
(\cite{Turland75,Waggett77,Readhead78,Perley79,Bridle84}). The
existence of X-ray emission from the nearest, brightest jets in Cen A
(\cite{Feigelson81}), 3C273 (\cite{Willingale81}), and M87
(\cite{Schreier82}) resulted from observations by the \emph{ROSAT} and
\emph{Einstein} X-ray telescopes, each with about 5\arcsec\ angular
resolution.  The \axaf\ X-ray observatory (\cite{Weisskopf02,
  Weisskopf03, Schwartz14}) with its 0\farcs5 resolution telescope
gives a 100-fold increase in 2-dimensional imaging capability. This
has led to the discovery of X-ray jets in a wide variety of
astronomical systems (\cite{Schwartz10}), and in particular has
exploded the study of X-ray emission from extra-galactic radio jets.

\subsection{Importance of jets}

In their 1984 review, \cite{Begelman84} wrote ``\dots the concept of a
jet is crucial to understanding all active nuclei, \dots .'' Jets can
carry significant amounts of energy, and since that power is not
subject to the Eddington radiation limit jets may allow
super-Eddington accretion rates. Such accretion may be relevant to the
growth of super-massive black holes in the early universe. High energy
observations using the \axaf\ Observatory revealed the effects of jets
on the gas filling clusters of galaxies \cite{Fabian00}. This solved a
long standing problem that the cooling time of gas in clusters of
galaxies was much less than the Hubble time, implying that the cluster
gas should collapse catastrophically. Jets on parsec scales in the
nuclei of galaxies explain the blazar phenomena of rapid variability
and apparent superluminal expansion. It is now known from direct
imaging that regions within X-ray jets may be variable even 10's of
kpc from the black hole (\cite{Marshall10,Hardcastle16}). From
gravitational lensing observations (\cite{Barnacka18}) of $\gamma$-ray
blazars it was found that variability could originate from regions
many kpc from the black hole (\cite{Barnacka15}) and that $\gamma$-ray
flares occur at locations distinct from the radio core
(\cite{Barnacka16}). These X-ray and $\gamma$-ray variability cases
constrain the mechanisms of particle acceleration.

\subsection{Outline of this review}

This review considers non-thermal jets from extra-galactic sources, and
will emphasize X-ray observations of powerful quasars. In the case of
FR-I radio sources, the X-ray jet emission can generally be interpreted as
an extension of the radio synchrotron spectrum
(\cite{Worrall05,Harris06,Harris07,Worrall09}). The subject of this
review will be FR-II quasars, in which case an optical flux or upper
limit generally shows that the X-rays can not be an extension of the
radio synchrotron emission. In section 3 we will see how this gives
information on the relativistic parameters of the jet, and/or on
multiple distinct populations of relativistic electrons.

\section{Application of the minimum energy assumption}

The intensity of synchrotron radiation from a power law distribution
of relativistic electrons, dN/d$\gamma$=$\kappa \gamma^{-(2\alpha+1)}$
is proportional to the product of $\kappa$ and the magnetic field
strength B. Here, $\alpha$ is the energy index of the observed
radiation, $\gamma m c^2$ is the electron energy, and the spectrum is
usually assumed to extend from a minimum $\gamma_1$ to a maximum
$\gamma_2$. From the radio synchrotron flux density alone, one cannot
determine either the magnetic field strength or the relativistic
particle density. Another relation between B and $\kappa$ can be
obtained by assuming minimum total energy in particles and fields,
which is nearly equivalent to assuming equipartition of energy between
those two channels (\cite{Burbidge56}). This assumption is now widely
used for the interpretation of sources emitting synchrotron radiation.

\cite{Miley80} has previously discussed the assumptions necessary to apply the
minimum energy condition. We update that discussion by considering the
relativistic particle spectrum to extend from  $\gamma_1$ to
$\gamma_2$ instead of fine tuning the  $\gamma$ to extend from the limits of
observed radio frequencies, and by considering the application to
X-ray jet measurements. This picture allows the equation for the
minimum energy magnetic field strength to be written in the form given
by \cite{Worrall09}:

\boldmath
$B_{min}=\textcolor{red}{f_{min}}[G(\textcolor{darkgreen}{\alpha})(1+\textcolor{red}{k})\textcolor{darkgreen}{L_{\nu}\nu^{\alpha}}(\textcolor{blue}{\gamma_1}^{1-2\textcolor{darkgreen}{\alpha}}-\textcolor{blue}{\gamma_2}^{1-2\textcolor{darkgreen}{\alpha}})/(\textcolor{red}{\phi}
\textcolor{darkgreen}{l} \textcolor{blue}{r} \textcolor{red}{t})]^{1/(
  \textcolor{darkgreen}{\alpha}+3)}$,\\
where G($\alpha$) is a combination of physical constants and functions
of $\alpha$.

In the on-line version we have color-coded the symbols as follows:
green symbols for quantities which can be measured directly, namely
the spectral index $\alpha$, the luminosity density $L_{\nu}$ at
frequency $\nu$, and the length of the jet element $l$. In turn, the
luminosity is determined from the flux density and the length is
determined from the angular extent by using the measured redshift and
a cosmological model. Here we use H$_0$=67.3 km s$^{-1}$ Mpc$^{-1}$
and $\Omega_m$=0.315 from the \cite{Planck17} results, in a flat
universe. Blue symbols are not directly measured, but have some
observational limits. These include the extremes of the power law
electron spectrum, $\gamma_1$ and $\gamma_2$, and the width of the jet
$r$. We note that $\gamma_2$ has a negligible effect on the
calculation, since it is typically orders of magnitude larger than
$\gamma_1$.  Red symbols are based on intuitive or simplifying
assumptions, such as the ratio, $k$, of energy in relativistic protons
to the energy in electrons (including positrons), the filling factor,
$\phi$, of particles and fields in the jet, the line of sight
thickness, $t$, through the jet, and the assumption, $f_{min}$ that
the magnetic field strength corresponds exactly to the minimum energy
condition. Here we will take the values $k=\phi=f_{min}=1$ and 
$t$=$r$. In addition to assuming minimum energy, we must transform
observed quantities to the jet rest frame using the Doppler factor
$\delta=1/(\Gamma(1-\beta \cos(\theta)))$ where $\theta$ is the angle
of the jet to our line of sight and $\Gamma$ is the Lorentz factor of
the jet relative to the CMB frame. The above discussion shows that we
must consider possible magnetic field strength uncertainty of a factor
of 2 or 3 in any modeling.

The \emph{Lynx} X-ray observatory
(\cite{Gaskin17,Gaskin18,Ozel18,Vikhlinin18}),  
which is being studied for submission to the 2020 Decadal Committee
for Astronomy and Astrophysics, would have 0\farcs5 angular resolution
and 2 m$^2$ effective area. The \emph{Lynx} capabilities will allow
significant improvements in the determination of some of the quantities needed to
calculate $B_{min}$ for X-ray jets. Currently, the modest photon
statistics from \axaf\ observations allow only an upper limit of
$\approx$ 0\farcs5 for the width, $r$, and assumed thickness, $t$, of
X-ray jets. With the 30-fold increase in collecting area, \emph{Lynx}
should allow measurements of widths down to 0\farcs1, or of order 700
pc even for the most distant jets. At 200---300 eV, the \emph{Lynx}
throughput will be 100 times that of \axaf . This should allow direct
determination of $\gamma_1$ via measurement of the soft X-ray
turn-over at those energies. With \axaf\ this measurement could only
be applied to PKS0637-752 (\cite{Mueller09}) due to the build-up of
contamination on the filter of the ACIS camera. The improved
statistics will also allow precise measurement of the jet X-ray spectral
index $\alpha$. The iC/CMB mechanism assumes this is the same index as
the GHz spectrum, although it could be flatter if the radiative
lifetime of the GHz emitting electrons is comparable to or less than
the age of the jet.

\section{The iC/CMB interpretation\label{icCmb}}

The X-ray emission of the luminous jet in PKS 0637-752 could not be
explained by reasonable models of synchrotron, inverse Compton, or
thermal mechanisms (\cite{Schwartz00,Chartas00}). \cite{Tavecchio00}
and \cite{Celotti01} provided the insight of invoking relativistic
bulk motion of the jet with Lorentz factor $\Gamma$, and using the
result from \cite{Dermer94} that this increased the energy density of
the cosmic microwave background in the rest frame of the jet by the
factor $\Gamma^2$. Subsequently the iC/CMB model has been widely used
to interpret the X-ray emission from the jets of powerful quasars
(\cite{Siemiginowska02,Sambruna02,Sambruna04,Sambruna06, Marshall05,
  Schwartz05, Marshall11, Marshall18, Schwartz06, Schwartz06b,
  Worrall09,Perlman11,Massaro11}).

\begin{figure}[t]
% \vspace*{-2.0 cm}
\begin{center}
 \includegraphics[width=5.in]{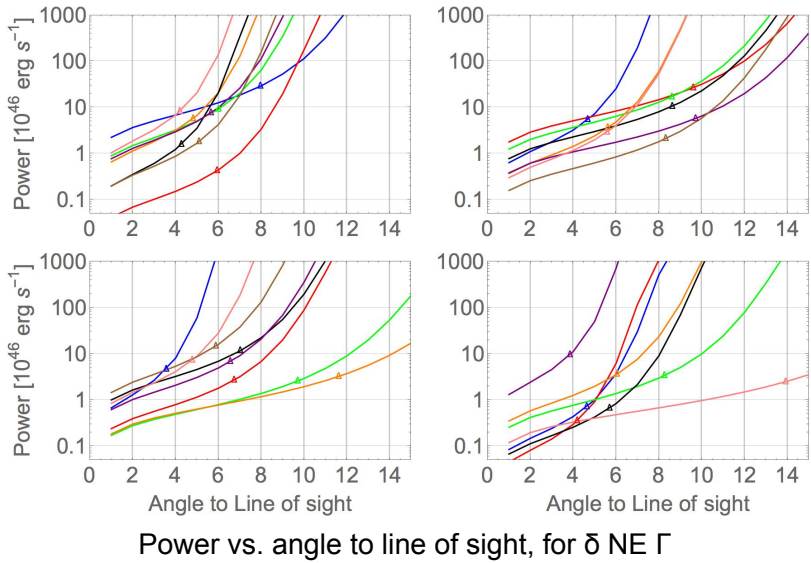} 
% \vspace*{-1.0 cm}
 \caption{Each curve in one of the panels gives what would be the
   enthalpy flow in one of the 31 detected jets, calculated as a
   function of the unknown angle of that jet to our line of sight, and
   using the iC/CMB model. The triangle on each curve marks the point
   for which $\Gamma = \delta$, and gives reasonable values for the
   distribution of parameters of the jet sample.}
   \label{fig:powerVSangle}
\end{center}
\end{figure}

Arguments supporting the iC/CMB model were originally based on the
similarity of the X-ray and the radio jet profiles over extended
angular distances, with the interpretation that they therefore
originated from a single, broad relativistic electron population.
(\cite{Schwartz00,Schwartz05b,Harris17}), Further evidence is suggested
by the  termination of X-ray emission just where the radio
jet goes through a large change of direction
(\cite{Schwartz03,Schwartz10}). This is naturally explained by the
X-ray to radio ratio of $\delta^{1+\alpha}$ discussed by
\cite{Dermer95}. However, powerful evidence against the iC/CMB mechanism in the
case of PKS 0637-752 was presented by \cite{Meyer15}, and
\cite{Meyer17}, 
while \cite{Breiding17} presented similar evidence for 3 jets for
which the iC/CMB model had not been indicated. They showed
\emph{Fermi} upper limits to GeV $\gamma$-ray emission was       
below the level expected \emph{if} the electron spectrum producing the
GHz radio emission also produced the jet  emission observed in the
\emph{ALMA} and IR/optical regions. 
In any event, there must be two distinct populations of relativistic
electrons, and the X-rays could possibly arise from iC/CMB if the
second population produced the \emph{ALMA} and optical emission. The
existence of two populations immediately shows that the assumption
on the filling factor, $\phi=1$, is not correct.

In this review we will continue to interpret the X-ray emission as
iC/CMB, which we note involves no new parameters for these relativistic
jets. In particular, we discuss the results from the systematic survey
of \cite{Marshall05,Marshall11,Marshall18} and previously presented by
\cite{Schwartz15}. Figure~\ref{fig:powerVSangle} shows the deduced
enthalpy power (\cite{Bicknell94}) carried by the 31 detected jets,
parameterized as a function of the unknown angle to our line of sight.
Qualitatively it is apparent that very small angles imply a very large
population of unrecognized sources which lie at larger angles. Large
angles, greater than 6\degs\ to 12\degs\ for most of the objects,
imply unrealistically large powers; namely, greater than 10$^{49}$ ergs
s$^{-1}$. An assumption is often made that $\Gamma=\delta$. This
corresponds to the jet being at the largest possible angle to our line
of sight for the given value of $\delta$. Those points are indicated
by the triangles on each curve, and
we expect they give a reasonable estimate of the distribution of
results from the sample.

\begin{figure}[t]
% \vspace*{-2.0 cm}
\begin{center}
 \includegraphics[width=5.5in]{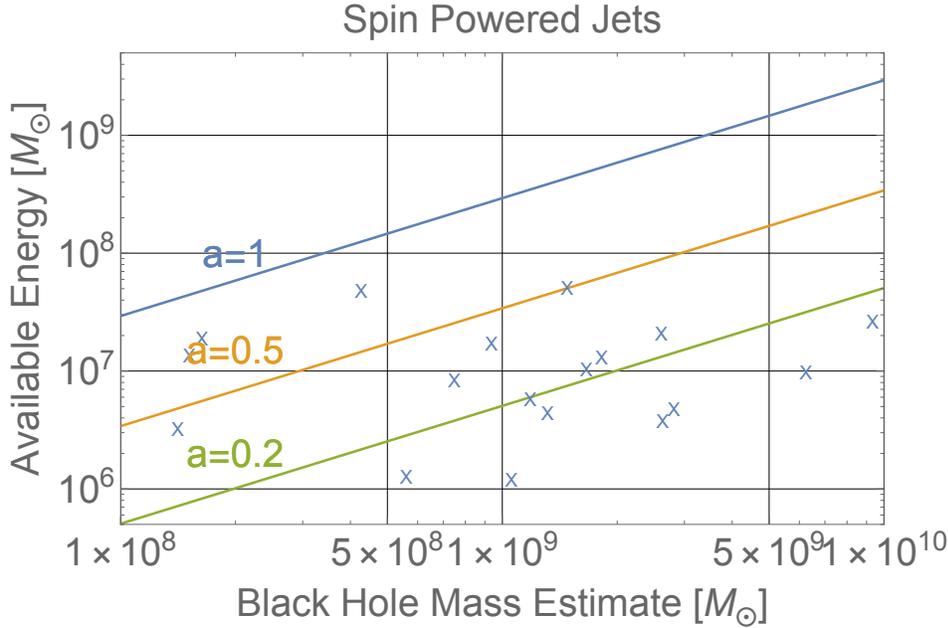} 
% \vspace*{-1.0 cm}
 \caption{The solid lines show the available rotational energy, in
   terms of rest mass, which can be recovered by optimal spin-down of
   a black hole as a function of initial mass. The different spin
   parameters a=1,0.3,0.2 times M, respectively 
allow 29\%, 1\%, and 0.5\% of the mass-energy of the black hole to be
recovered in principle. The x's plot the amount of energy required to sustain the
   power of the jets for 10$^7$ years. The black hole mass estimate
   here is taken as a median value of estimates in the literature
   tabulated by \cite{Shen11}, and by \cite{Xiong14} and   references
   therein. Even black holes with only 0.2 to 0.3 of the maximum
   possible angular momentum, can power the jets we measured for
   millions of years.}
   \label{fig:spinPowered}
\end{center}
\end{figure}

\section{Connection to the super-massive black hole}

The rotational energy, E$_r$ of a Kerr black hole manifests as a
contribution E$_r$/c$^2$=M$_r$ to the total mass of the black
hole. The relation is non-linear, with the fraction 
\begin{center}
M$_r$/M =1- $\sqrt{0.5(1+\sqrt(1-a^2/M^2))}$ 
\end{center}
being available in
principle to be expelled. Here, $a$ is the spin parameter in units of
the total black hole mass $M$. The solid lines in
Figure~\ref{fig:spinPowered} show the available mass that can, in
principle, be released as a function of the total black hole mass. The
$x's$ plot the mass-energy equivalent required  to power the
jets for 10$^7$ years, as determined from the $\Gamma=\delta$ point in
Figure~\ref{fig:powerVSangle}. We plot those masses against the estimated
mass of the given black hole. That black hole mass is taken from the
median of values given in the literature for 18 objects where it is
available. We see that maximally spinning black holes, $a \approx 1$
can provide more than the required energy, and super-massive black
holes with spin parameters even as low as 0.3 could in principle power the
observed jets for millions of years. Of course, this consideration
does not address the dynamics of actually extracting such energy in a
collimated flow.

\begin{figure}[h]
% \vspace*{-2.0 cm}
\begin{center}
  \includegraphics[width=4.5in]{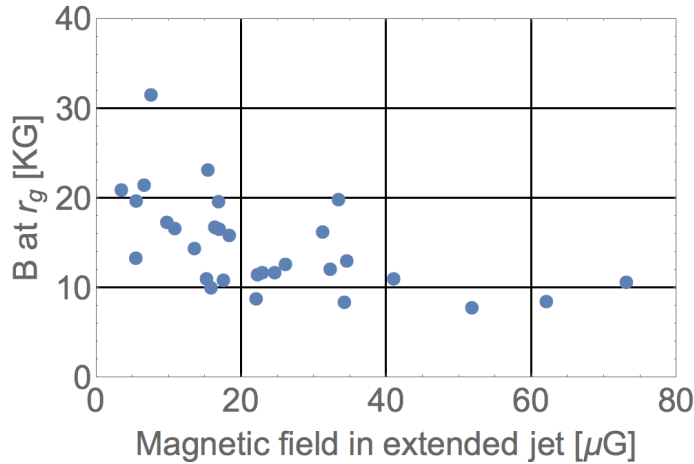} 
% \vspace*{-1.0 cm}
 \caption{Ordinate: Magnetic field strength at the gravitational
   radius such that the Poynting vector equals the power deduced for
   the kpc-scale jet. Abscissa: Mean magnetic field strength in the
   rest frame of the kpc scale jet, according to the iC/CMB model.
   Points are the 31 jets detected in the survey discussed in
   section~\ref{icCmb}. (In view of
numerous uncertainties, we do not consider the possible
anti-correlation to be significant.)}
   \label{fig:BvsB}
\end{center}
\end{figure}

Numerical calculations of the magnetically arrested disk (MAD) model
(\cite{Narayan03,Igumenshchev08,Sadowski14,Tchekhovskoy11,Zamaninasab14})
for a rapidly spinning black hole have had some success showing that
jets can be formed, and can extract more than the potential energy of
the accreting matter. 
A magnetic field is advected in with the accreting matter, and is
compressed and amplified until its pressure balances the gravitation
pull (\cite{Narayan03}). This results in chaotic variability of the accretion, and the
rapidly spinning black hole wraps the magnetic field lines into a
tight spiral about the spin axis. Observations of rotation measure
gradients across pc-scale radio jets provides evidence for such field
geometry (\cite{Gabuzda14, Gabuzda15, Gabuzda17}). The magnetic
pressure then causes ejection along the spin axis. If we assume that
the initial ejection right at the gravitational radius of an extreme
Kerr black hole is a pure Poynting flux, then we can obtain a lower
limit to that initial magnetic field by assuming 100\% efficiency for
converting that Poynting flux into the energy flux deduced for the
kpc-scale jet. Such initial field strengths are shown in
Figure~\ref{fig:BvsB}. We have taken the mass from
fundamental plane relation of \cite{Gultekin09} to deduce the
gravitational radius $r_g$ of each quasar.

\section{Summary}

We use Chandra X-ray observations to estimate the power of quasar jets, by
observing the jet itself. We tie this to the central black hole mass
on an individual object basis. The rotational energy of super-massive
black holes can power these quasar jets, even with spin parameters as
low as a/M=0.2, for lifetimes longer than millions of years. If the power
we observe originates as a pure Poynting flux, we derive initial
magnetic field strengths of a few 10's of kilo-Gauss. For models of
magnetically arrested disks  this inferred magnetic flux is
of the order of magnitude of predictions, for Eddington limited
accretion onto maximally spinning super-massive black holes.

\end{document}